\documentclass[conference]{IEEEtran}
\IEEEoverridecommandlockouts
\usepackage{cite}
\usepackage{amsmath,amssymb,amsfonts,stackrel,eqnarray,}
\usepackage{algorithmic}
\usepackage{graphicx}
\usepackage{textcomp}
\usepackage{xcolor}
\usepackage{multirow}
\usepackage{url}

\def\BibTeX{{\rm B\kern-.05em{\sc i\kern-.025em b}\kern-.08em
    T\kern-.1667em\lower.7ex\hbox{E}\kern-.125emX}}

\begin{document}

\title{Coded Backscattering Communication with LTE Pilots as Ambient Signal}

\author{\IEEEauthorblockN{Jingyi Liao\IEEEauthorrefmark{1}, Kalle Ruttik\IEEEauthorrefmark{1}, Riku J\"{a}ntti\IEEEauthorrefmark{1}, and Phan-Huy Dinh-Thuy\IEEEauthorrefmark{2}}

\IEEEauthorblockA{\IEEEauthorrefmark{1}Department of Communications and Networking,
Aalto University,
02150 Espoo, Finland}
Email: \{jingyi.liao, kalle.ruttik, riku.jantti\}@aalto.fi
 
\IEEEauthorblockA{\IEEEauthorrefmark{2}Networks, Orange Innovation, Chatillon, France}
Email: dinhthuy.phanhuy@orange.com
}
\maketitle

\begin{abstract}
The 3GPP has recently conducted a study on the Ambient Internet of Things (AIoT), with a particular emphasis on examining backscatter communications as one of the primary techniques under consideration. Previous investigations into Ambient Backscatter Communications (AmBC) within the long term evolution (LTE) downlink have shown that it is feasible to utilize the user equipment channel estimator as a receiver for demodulating frequency shift keyed (FSK) messages transmitted by the backscatter devices. In practical deployment scenarios, the backscattered link often experiences a low signal-to-noise ratio, leading to subpar bit error rate (BER) performance in the case of uncoded transmissions.
In this paper, we propose the adoption of the same convolutional coding methodology for backscatter links that is already employed for LTE downlink control signals. This approach facilitates the reuse of identical demodulation functions at the modem for both control signals and backscattered AIoT messages. To assess the performance of the proposed scheme, we conducted experiments utilizing real LTE downlink signals generated by a mobile operator within an office environment.
When compared to uncoded FSK, convolutional channel coding delivers a notable gain of approximately 6 dB at a BER of $10^{-3}$. Consequently, the AmBC system demonstrates a high level of reliability, achieving a BER of $10^{-3}$ at a Signal-to-Noise Ratio (SNR) of 5 dB.

\end{abstract}


\section{Introduction}
The Ambient Internet of Things (AIoT) ~\cite{3GPPAIoTSI} presents a demand for communication solutions tailored to Zero Energy Devices (ZEDs). AIoT relies on harvesting environmental energy to meet its power requirements.
The 6G flagship project Hexa-X II ~\cite{D52}, in continuity of Hexa-X-I ~\cite{HX}, is committed to designing such zero-energy communication solutions in general, and to developing and trialing on the field one particularly promising technique called Ambient Backscatter Communication (AmBC)~\cite{liu2013ambient}. In the AmBC approach, backscatter devices (BDs) do not generate their own signals; instead, they rely on ambient radio frequency (RF) signals such as television broadcast and cellular signals for illumination. Information in AmBC is conveyed through modulation of the propagation channel, as BDs either reflect or absorb ambient RF signals. 

The primary advantage of utilizing backscatter communications in AIoT resides in its exceptional energy efficiency. When compared to active transmitters, backscatter devices demonstrate a remarkable reduction in energy consumption, approximately 1000 times less, as discussed in~\cite{talla2017lora}. This substantial reduction in energy consumption lays the foundation for the deployment of ZED, which can operate without batteries. Such devices have valuable practical applications, including asset tracking~\cite{phan2021ambient} and proximity-based localization~\cite{localisation}.

A significant challenge in AmBC is dealing with direct path interference (DPI) caused by the higher power of the direct path signal compared to the scattered signal. Another challenge, especially in the case of ambient cellular signals, is the intermittent and bursty nature of the signals. 

In long term evolutio(LTE) downlink, the receiver at the user equipment (UE) first synchronizes with the Base Station (BS, a.k.a eNB) using primary and secondary synchronization signals (PSS and SSS) and then employs the downlink Cell-Specific Reference Signals (CRS) to estimate the channel~\cite{3gpp2010evolved}. In~\cite{Ruttik2022, Liao2023jrfid}, an in-band ambient Frequency Shift Keying (FSK) backscatter communication system was proposed for LTE downlink, leveraging the UE receiver for demodulating the backscatter-generated FSK modulated messages. There reason for FSK modulation scheme is that FSK is a better modulation scheme than BPSK in AmBC \cite{8474355}. 

The LTE channel estimator is engineered to withstand significant Doppler shifts, typically encountered in high-mobility scenarios. However, it's essential to emphasize that in the majority of indoor and pedestrian deployments, the Doppler shift remains considerably below the maximum threshold. The proposed in-band LTE backscatter system effectively harnesses this excess bandwidth to facilitate backscatter communication. The BD employs FSK to introduce an artificial Doppler shift, which is larger than the natural Doppler within the channel yet smaller than the maximum Doppler tolerated by the channel estimator. This strategic choice enables the receiver to discern BD-modulated messages from the natural multi-path components in the Doppler frequency domain, thereby resolving the DPI issue. Additionally, CRS symbols are transmitted with fixed and often boosted power, even during cell idle periods, enabling the receiver to independently demodulate BD messages without dependence on LTE downlink traffic.

In prior studies concerning in-band LTE backscatter systems, the primary emphasis has been on uncoded modulation as the predominant operational mode for the BD. Empirical measurements~\cite{10317672} have unveiled that the system's bit-error rate performance is poor in many practical scenarios. In response to this challenge, the integration of forward error correction (FEC) coding within the BD is proposed in this paper. To optimize the utilization of existing receiver components within the UE, the adoption of the same convolutional code for BD messages as is conventionally employed for downlink control messages is recommended. We validate the proposed system with over the air measurements by using ambient signal from commercial BS and processing the BD signal as it would be if an LTE UE would be the BD signal reader.

The rest of paper is organized as follows: Sec. \ref{se:System} discusses AmBC model and how BD signal is visible on UE side. Then, Sec. \ref{se:receiver} outlines the AmBC receiver signal processing flows. Next, Sec. \ref{se:experimentation} shows the experiment measurements under commercial LTE BS. Finally, Sec. \ref{se:conclusions} draws conclusions of this study.

\section{System model\label{se:System}}
The top-level of in-band LTE backscattering based AmBC system consists three parts, BS, UE and BD.
BS continuously boardcasts LTE downlink pilots. 
Those ubiquitous continuous pilots enlighten BD. UE always estimates the channel base on pilot CRS signals.
BD modulates the channel blindly, without know any ambient information. Once UE is close to BD, it can recognize the channel changes. 
In this way, information is sent from BD to UE.

As Fig. \ref{fig:setup} illustrates. BS sends CRS to UE. BD reflects or absorbs the incident RF ambient signal, toggles states with different frequency and generates artificial Doppler frequencies. Those channel changes are picked up by UE, i.e. UE as AmBC receiver demodulates BD information based on the channel estimates. 


\subsection{Channel Model}

In the context of the LTE system, the variables $y[n]$ and $x[n]$ represent samples corresponding to transmitted and received LTE signals. The complex channel from the BS to the UE is denoted as $h_r$. We note that the AmBC system operates independently of the LTE system. Specifically, BD is not synchronized with LTE nor it is aware of the downlink transmissions. Within the framework of AmBC, the backscatter link comprises two distinct components: the channel from the BS to the BD, denoted as $h_t$, and the channel from the BD to the UE, represented by $h_b$.

The ambient backscatter signal model is as \cite[Eq. 2.11]{devineni2021ambient}
\begin{equation}
    y[n] =  (h_r+ M h_b b h_t)x[n]+w[n],
\end{equation}
where $b$ corresponds to off or on state for BD at sample time $n$ and is either 0 or 1, $M$ relates to scattered signal power strength from BD, and $w(n)$ is i.i.d. additive white Gaussian noise.
Limited by the hardware design, BD flips between two states, reflecting incident signal or absorbing incident signal. Therefore in such a system, a FSK is approximated by square-wave FSK.

The channel estimation $\hat{h}[m]$ in UE recognizes the changes in propagation environment. According to LTE specification~\cite{3gpp2010evolved}, channel is estimated base on CRS. Those channel estimates include the artificial Doppler caused by BD square wave FSK. BD generates square wave FSK by toggling on and off states in two different speed. The LTE $m$-th OFDM symbol estimates channel as~\cite{Liao2023jrfid}
\begin{align}
    |\hat{h}[m]|^2 &\approx |h_r[m]|^2 + \nonumber \\
    &\left[ |M h_b[m] h_t[m]|^2 + 2\text{Re}(M h_b[m] h_t[m]h_r^*[m]) \right] b[m].
\end{align}
Study~\cite{Liao2023jrfid} shows a method that the AmBC receiver extracts channel change from pilot channel estimates. 

\section{Backscatter receiver \label{se:receiver}}
\begin{figure}[t!]
\centering
    \includegraphics[width=\columnwidth]{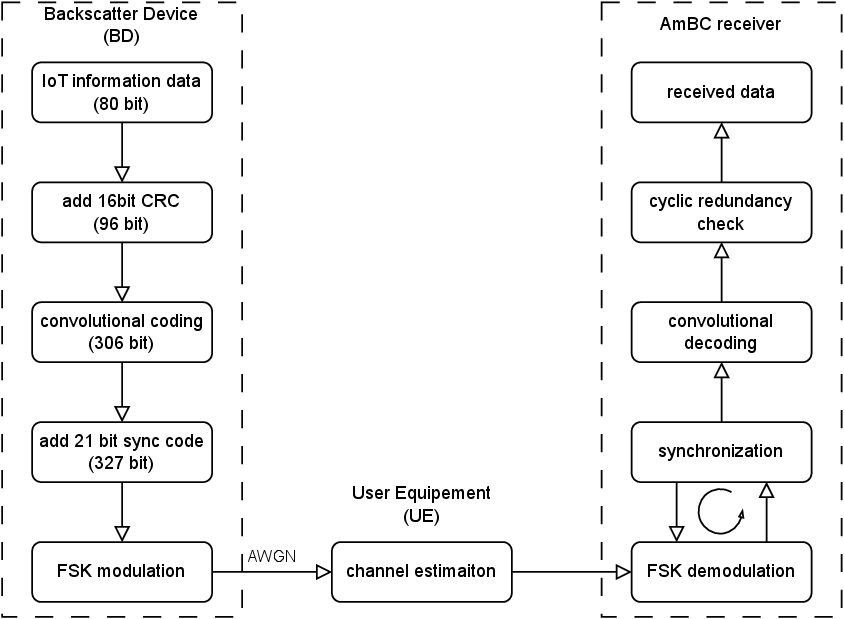}
  \caption{Flow chart of the proposed backscatter receiver.}
  \label{fig:Flowchart}
   \vskip -12pt
\end{figure}

The AmBC receiver efficiently repurposes the LTE Rx channel estimation, convolutional coding algorithm, and cyclic redundancy check (CRC). These components are intentionally designed with identical parameters to minimize receiver cost allowing the existing software components in the modem to be reused.
The block diagram of the proposed backscatter receiver is illustrated in Fig.~\ref{fig:Flowchart}. The frame packet structure is organized as Fig. \ref{fig:dataFormat}. In this section, we explain the FSK demodulator and convolutional decoder in more details.

\subsection{Symbol and frame synchronization}
\begin{figure}[t!]
\centering
    \includegraphics[width=\columnwidth]{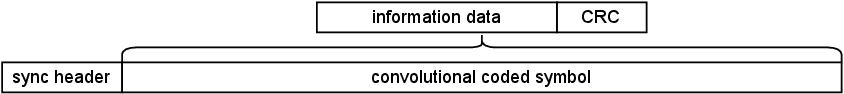}
  \caption{A frame packet structure.}
  \label{fig:dataFormat}
   \vskip -12pt
\end{figure}

As previously discussed, the BD operates without synchronization to LTE transmissions. To enable the receiver to detect the start of the BD frame, a synchronization header is employed. The frame incorporates a synchronization header comprising three 7-bit Barker codes and 306 convolutional coded data bits, as illustrated in Fig. \ref{fig:dataFormat}. The complete packet comprises a total of 327 symbols. The synchronization preamble is comprised of three sequences of 7-bit Barker code. Specifically, two standard 7-bit Barker code sequences, denoted as '1110010', are concatenated with an inverted 7-bit Barker code sequence, '0001101', effectively anchoring the header of a frame. Our measurements confirm that the sidelobe auto-correlation of these three 7-bit Barker codes is sufficiently low for effective synchronization.

Barker code synchronization is crucial because it serves as both the reference for the starting symbol of a frame and the reference for the starting sample of FSK symbols. The starting sample is determined by locating the peak of the auto-correlation between the received signal and the synchronization preamble. The header remains continuously synchronized within a frame packet time window. If there were any sample shifting in header synchronization, FSK demodulation would introduce high noise levels. The synchronization process involves a collaborative effort between Barker code synchronization and FSK demodulation, as depicted in Fig.~\ref{fig:Flowchart} with the circular arrow. The sequence diagram for the synchronization loop software threads can be found in~\cite[Fig. 2]{10.1145/3581791.3597285}, and the synchronization code is available as open source on GitHub \cite{GitHub}.

\subsection{Square-wave FSK}
We desing FSK to have higher frequency than natural Doppler frequency of a slow moving user.
The square-wave FSK is demodulated by noncoherent detector after a low pass filter. This low pass filter removes the natural Doppler effect and frequency offset due to the clock drifting between transmitter and receiver. 
There are upper bound of FSK symbol frequency, Nyquist frequency. Each FSK symbol is oversampled by channel estimation speed. We select FSK frequencies to be less than 1 kHz, since in LTE the channel is estimated with 4 kHz, the pilots occur about every 250 $\mu$s. However the pilots are inserted slightly non-uniformly ~\cite{3gpp2010evolved} and therefore there is 1 kHz harmonic 
\cite{Liao2023jrfid}. Hereby, the FSK frequency constraints to a narrow band, about several hundreds Hertz.
\par
The noncoherent detection corresponds power detector of the channel estimates.
After obtaining symbol synchronization, the FSK demodulation receiver correlates the oversampled symbol sequence on several different lags. The maximum correlation peak provides us the start time of the frame.
\par
Limited by the narrow FSK band, bit rate of AmBC $1/T_s$ is low, 25 bit/s in this case. This data rate is selected because detector needs longer symbol length for oversampling, to collect enough energy for symbol decision.

\subsection{Channel coding and CRC}
Due to the absence of a buffer in this software implementation, the total frame length cannot be extended significantly. Each frame transmitted by the BD contains only an 80-bit information sequence. Additionally, a 16-bit CRC (CRC-CCITT) code is appended to the end of the information data to facilitate error bit detection and correction.

The CRC-CCITT cyclic generator polynomial, denoted as $g_\text{CRC16}$, is expressed as follows:
\begin{align}
g_\text{CRC16}(D) = D^{16}+D^{12}+D^{5}+1.
\end{align}
This polynomial operates within the finite field $GF(2)$ in binary notation\cite{3gppConvEnc}. The CRC tail is generated based on the remainder of the information bit sequence divided by the hexadecimal number 0x11021. For a more detailed description of the CRC generator, please refer to~\cite{10.5555/197029}.

Within the LTE technical specification, there exist four types of cyclic generator polynomials employed to generate parity bits: CRC-8-WCDMA, CRC-CCITT, CRC-24-Radix-64, and CRC-24-WCDMA~\cite{3gppConvEnc}. Given that AmBC is susceptible to the effects of multipath fading, errors occurring over several consecutive symbols, often referred to as burst errors, are common. The 8-bit CRC, while effective, can detect only 98.44\% of all error bursts with a length of 8 bits, which is deemed too short for handling the data packet in this context. On the other hand, the 24-bit CRC is overly long for this data packet, resulting in a reduction in effective information baud. In light of these considerations, the 16-bit CRC-CCITT emerges as an optimal choice, as it effectively detects 99.997\% of error bursts with a length greater than 17 bits and captures all error bursts with a length of fewer than 16 bits, striking a perfect trade-off \cite{10.5555/197029}. On the AmBC receiver side, received symbols undergo CRC checking after convolutional soft-decision Viterbi decoding.

There are many channel coding candidates, such as LDPC, Turbo, and polar coding. LTE specifies convolutional code for control channel and turbo code for data channels. Our practical backscattering implementation has a short 80-bit data packet. For such a short code convolutional coding outperforms other coding options, as demonstrated in ~\cite{7998249,7938068,7848804}. LTE control channel uses a tail-biting convolutional code with a 1/3 coding rate and a constraint length of 7~\cite{3gppConvEnc}. This straightforward convolutional structure is easy to implement in a low cost BD device. 

\begin{figure}[t!]
\centering
    \includegraphics[width=\columnwidth]{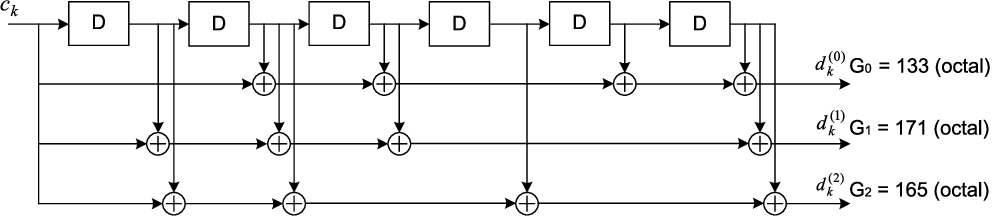}
  \caption{An LTE rate 1/3 tail-biting convolutional coding encoder. \cite{3gppConvEnc}}
  \label{fig:ccEncoder}
   \vskip -12pt
\end{figure}




In Fig. \ref{fig:ccEncoder}, the encoder takes an information data stream as input, denoted as $c_k$, and produces output streams at digital clock sample time $k$, represented as $d_k^{(0)}$, $d_k^{(1)}$, and $d_k^{(2)}$.
The output stream at sample time $k$ is
\begin{align*}
    \begin{bmatrix}
         d_k^{(0)} \\ d_k^{(1)} \\ d_k^{(2)}
    \end{bmatrix}
    =
    \begin{bmatrix}
         1&0&1&1&0&1&1\\
         1&1&1&1&0&0&1\\
         1&1&1&0&1&0&1\\
    \end{bmatrix}
    \begin{bmatrix}
        c_k\\c_{k-1}\\\vdots\\c_{k-6}
    \end{bmatrix}
    .
\end{align*}
This code generator is optimized, reaching upper bound free distance $d_\text{free}$ for coding rate $R_c=1/3$, $d_\text{free}$=15 \cite[Tab. 8.3-2]{massoud2007digital}.
After channel coding, the encoded bits is three times than bit number of information sequence and 16-bit CRC, padding with 6 zeros tail. Considering 21-bit synchronization header, the number of symbol in one frame reaches to 327 symbol in total, which spends about 13 seconds for only one frame.

For convolutional decoder, the soft-decision Viterbi decode algorithm is applied. 
However, convolutional coding is optimal under BPSK or QPSK. In this implementation we use FSK and that deteriorates the BER performance.

\section{Experimental results \label{se:experimentation}}
\begin{figure}[t!]
    \centering
    \includegraphics[width=0.75\columnwidth]{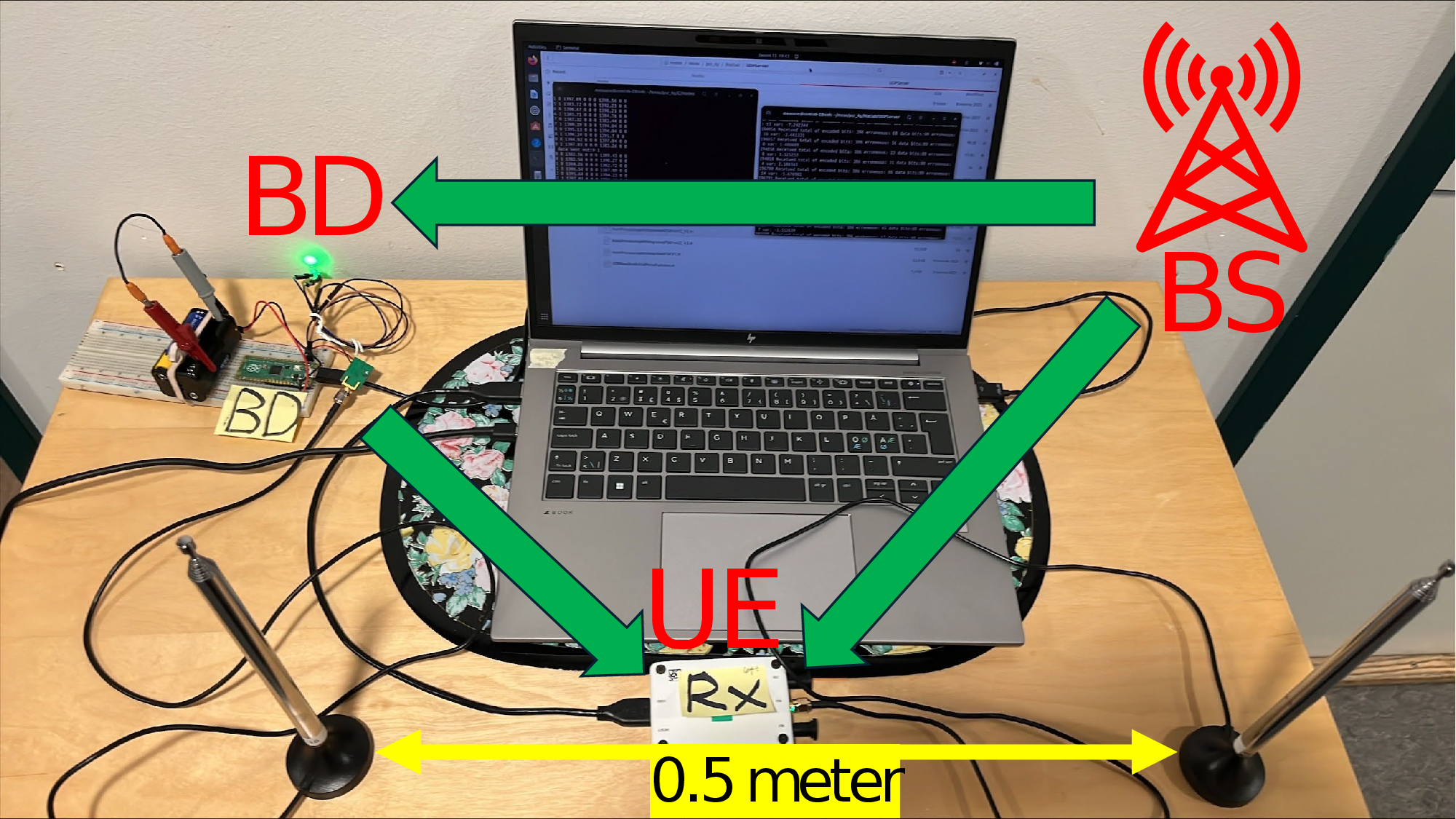}
    \caption{An illustration of the measurement setup.}
    \label{fig:setup}
\vskip -12pt
\end{figure}

To evaluate the feasibility of the proposed system, an experiment was conducted using real commercial LTE downlink signals within the indoor environment of the second floor of the Aalto University TUAS building in Espoo, Finland. Fig. \ref{fig:setup} illustrates the wireless LTE downlink measurement setup. Tab. \ref{tab:parameters} presents some essential parameters related to the BS, BD, and UE. The experiment uses a commercial BS, providing a reference signal received power (RSRP) of approximately -70 dBm at the measurement location. The distance between the BD and UE is half a meter.

The BD operates create its packets without any knowledge of the LTE downlink signals. The BD was a Rasperry Pi connected to a diode based backscatter module. The generated frame was 13.08 s. Each frame was followed by half a second sleep after which BD restarts transmission. 
This half-second silence serves the purpose of separating two consecutive frames, allowing the receiver to synchronize with the next frame using its synchronization header.

On the reader end, the UE equipment is emulated as a software-defined radio (SDR) with computer and B210 USRP frontend from National Instrument. 
This emulated UE synchronizes with the commercial LTE downlink, performs channel estimation, and subsequently transmits the channel estimates to a laptop via the User Datagram Protocol (UDP).


To counteract clock drifting in B210 
the frame length is deliberately kept short, and synchronization occurs frequently. This approach helps prevent the accumulation of clock errors, ensuring that symbols can be demodulated accurately within their correct symbol duration. Although there are many good channel coding for long packet, but that clock drifting prevents to design long packet.

The received channel estimates are post processed in MATLAB program, which serves as the AmBC receiver module depicted in Fig. \ref{fig:Flowchart}.

It's worth noting that, in comparison to mass-produced UE transceiver RF front-ends, the USRP is equipped with a 12-bit analog-to-digital converter (ADC) and is not using automatic gain control (AGC). To achieve an optimal balance between the dynamic range, quantization error, and signal clipping, the gain settings are selected by hand.

\begin{table}[t!]
\centering
\caption{Measurement setup parameters}
\resizebox{\columnwidth}{!}{
\begin{tabular}{l|c|c} \hline
   &Parameter &Value \\
   \hline
   \multirow{3}{*}{BS} & Commercial BS & LTE downlink \\ \cline{2-3} 
   & Carrier frequency & 806MHz \\ \cline{2-3} 
   & Bandwidth & 7.68 MHz \\ 
   \hline
   \multirow{7}{*}{BD}
   & Micro controller & Raspberry Pi nano \\ \cline{2-3} 
   & Antenna & RaTLSnake M6 telescopic antenna\\  \cline{2-3} 
   & Symbol duration & 40 ms \\ \cline{2-3} 
   & Synchronization & three 7-bit Braker code \\ \cline{2-3} 
   & Encoding & Convolutional coder, R=1/3 \\ \cline{2-3} 
   & Modulation scheme & FSK, $f_0=250$ Hz and $f_1=625$ Hz\\ \cline{2-3}
   & Sync header & Three 7-bit Barker code\\ 
   \hline
   \multirow{3}{*}{UE} & Device & NI USRP-B210 \\ \cline{2-3} 
   &Antenna & RaTLSnake M6 telescopic antenna\\ \cline{2-3} 
   &AD converter& 12 bits\\ 
   \hline
\end{tabular}}
\label{tab:parameters}
\end{table}

\begin{figure}[t!]
\centering
    \includegraphics[width=\columnwidth]{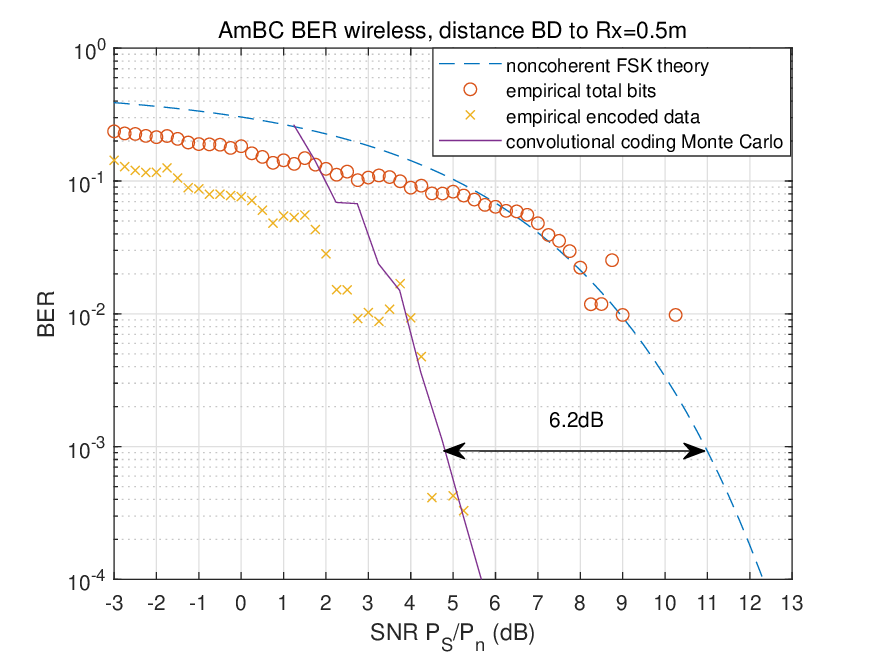}
  \caption{Measured SNR and BER in commerical LTE downlink environment. \label{fig:SNR_and_BER} }
  \label{fig:ReceivedSignal} 
\vskip -12pt
\end{figure}

The measuements we done by keeping measurement setup in Fig.~\ref{fig:setup} in location. 
The measurement campaign was carried out during two days. During that time the channel varies due to users movement in surrounding. During the campaign we collected 6050 packets. For each packet we recorded the signal-to-noise ratio (SNR) and amount of errors. 
In BER computation we split the SNR into 0.25 dB intervals. The final bit error rate (BER) is computed over all the packets that are in the same interval Fig.~\ref{fig:SNR_and_BER}.   


In Fig.~\ref{fig:SNR_and_BER} we give also uncoded and convolutional coded FSK theoretical performance curves.
The theoretical noncoherent FSK BER $P_b$ is \cite[Eq. 4.5-46]{massoud2007digital}
\begin{equation}
    P_b=\frac{1}{2}\exp\left(-\frac{\mathcal{E}_b}{2N_0}\right),
\end{equation}
where $\mathcal{E}_b$ energy per bit and $N_0$ is noise power density.
The convolutional coding reference curve is evaluated by Monte Carlo simulation. 
For a convolutional code we can estimate a coding gain as 
\begin{equation}
    \text{Coding gain} \leq 10 \log_{10}(R_c d_\text{free}).
\end{equation}
For the used convolutional code and BPSK modulation the coding gain upper bound is 6.99 dB \cite[Tab. 8.6-1]{massoud2007digital}.
In Fig.~\ref{fig:SNR_and_BER} the coding gain at BER $10^{-3}$ is about 6.28 dB. 

At high SNR the measured results align well with the theoretical curve.
At low SNR the measured performance is better than theory predicts. This discrepancy is explained by synchronization failure at low SNR. In our measurements we ignored packets that we could not synchronize to. That biased the measurements.  
Conversely, at high end of SNR the outliers are due to the lack of measured packet samples in this area. 


The measurement results demonstrate that AmBC is viable even under a 5 dB SNR, achieving a BER of less than $10^{-3}$. This represents a gain of at least 6 dB when compared to uncoded alone. Furthermore, the implemented AmBC receiver takes advantage of LTE UE receiver. 
It reuses the UE existing modules such as channel estimation, convolutional channel decoder, and CRC. 

\section{Conclusions \label{se:conclusions}}

In this paper, we have integrated convolutional channel coding and CRC into an AmBC system that relies on LTE downlink pilots. Our experiments conducted in a real commercial LTE environment validate the feasibility of AmBC with the proposed receiver.
The introduction of convolutional channel coding has led to a noteworthy 6 dB improvement in BER performance, particularly at a challenging SNR of 5 dB. The results consistently show that this AmBC system can achieve a BER of at least $10^{-3}$ under the specified 5 dB SNR conditions, which are relevant for a BD and UE separation distance of half a meter. In summary, our research suggests the potential applicability of our AmBC solution in the context of 6G ZED communication, representing a notable advancement in the field.

\section{Acknowledgements}
This work is in part supported by the European Project Hexa-X II under (grant 101095759) and Business Finland Project eMTC (Dnro 8028/31/2022). 


\bibliographystyle{IEEEtran}
\bibliography{References}

\end{document}